\documentclass[12pt]{iopart}
\usepackage{amsfonts}

\newtheorem{theorem}{Theorem}
\newtheorem{definition}[theorem]{Definition}

\newtheorem{corollary}[theorem]{Corollary}

\newcommand{\tbar}{\overline{T}}
\newcommand{\beq}{\begin{equation}}
\newcommand{\eeq}{\end{equation}}

\begin{document}

\title[Spectral Invariants and Vacuum Energy] 
{Some Subtleties in the Relationships among Heat Kernel
  Invariants, Eigenvalue 
Distributions, and Quantum Vacuum Energy
{\qquad\normalfont\small version of 28 November 2014} 
}
\author[S A Fulling and Y Yang]
{S A Fulling$^1$ and Yunyun Yang$^2$}
\address{$^1$ Departments of Mathematics and Physics,
  Texas A\&M University, College Station, TX 77843-3368 USA}
\address{$^2$ Department of Mathematics,
  Louisiana State University, Baton Rouge, LA 70803 USA}
\eads{\mailto{fulling@math.tamu.edu}, 
\mailto{yyang18@math.lsu.edu} }

\begin{abstract}
A common tool in Casimir physics (and many other areas)
is the asymptotic (high-frequency) expansion of eigenvalue
densities, employed as either input or output of calculations
of the asymptotic behavior of various Green functions.
Here we clarify some fine points and potentially confusing
aspects of the subject.
In particular, we show how recent observations of
Kolomeisky et al.\ [Phys.\ Rev.\ A \textbf{87} (2013) 042519]
fit into the established framework of the distributional 
asymptotics of spectral functions.

\end{abstract}

\pacs{02.30.Lt, 03.65.Sq, 11.10.Gh}
  \ams{34B27, 35P20, 46F10, 81T55}
%\submitto{\JPCM special issue on Casimir Physics}
\submitto\JPA

\maketitle

\section{Introduction}  \label{sec:intro}
The  density of eigenvalues of a differential operator
(such as a Hamiltonian), with the closely related subjects of 
semiclassical approximations and the 
asymptotic dependence of various Green functions on various
coordinates and parameters,
is a central tool in nuclear, atomic, and molecular physics
\cite{BaBl}, condensed-matter physics  \cite{BrBh}, 
general-relativistic quantum field theory \cite{Chr},
and quantum vacuum energy (Casimir physics) \cite{systemat}.
(These four references are merely illustrative.)
A key point is that a variety of differential equations 
(relativistic and nonrelativistic, classical and quantum)
are all associated with the same elliptic, spatial differential
operator; a variety of Green functions (heat, Schr\"odinger,
wave, \dots) are related to the same eigenvalue distribution and
hence to each other, and the study of one may yield valuable
insight into another. 
Nevertheless, the subject contains some complications and pitfalls
that are likely occasionally to rise up and confuse even those 
of us who think we have become experts in it.

Here is a synopsis of the paper in mathematical language:
Let $-H$ be the Laplacian on scalar 
functions in a compact
region in $\bf R^3$ with smooth Dirichlet boundary;
let $K(t)= {\rm Tr\,} e^{-tH}$ be the heat kernel trace,
$T(t)={\rm Tr\,} e^{-t\sqrt{H}}$ the trace of the cylinder
(Poisson) kernel, and $N(\omega^2)$ the eigenvalue 
counting function.
Loosely speaking, the small-$t$ asymptotics of $K$ and $T$
are in close correspondence with the averaged large-$\omega$
asymptotics of $N$, but there are some subtleties that
can be confusing.
(1) Nonnegative integer powers of $t$ in the expansion
of $K$ do not correspond to (negative) integer powers
of $\omega^2$ in the expansion of $dN/d(\omega^2)$
(even after the latter has been well defined by averaging).
Instead, they give rise to terms $\delta^{(n)}(\omega^2)$
in the {\it moment asymptotic expansion\/} of a distribution
(which is actually an expansion in a parameter, not in
$\omega$). 
(2) The expansion of $T$ contains additional,
nonlocal spectral invariants, which show up in 
$dN/d\omega  = 2\omega\,dN/d(\omega^2)$, filling in the
missing odd negative integer powers.  The first of these,
$O(t^0)$, or its electromagnetic analogue,
gives the Casimir energy in quantum field theory with 
idealized boundary conditions.
(3) Negative powers in $T$ physically represent
``divergences'' that must be explained or argued away.
The term of order $t^{-1}$ is particularly subtle because
it corresponds to the first ``moment'' term in $N$
(also to the topological (index) or Kac term, $O(t^0)$, in $K$).
Thus there is no $O(\omega^{-1})$ term in $dN/d\omega$,
so this term in $T$ has correctly been said to come from 
low-frequency oscillations of the eigenvalue density 
rather than high-frequency asymptotics;
however, that does not mean that it is one of the nonlocal
cylinder-kernel terms.  Also,  although a divergent
local energy density near the boundary does exist, 
because of an algebraic accident (unrelated to the 
``moment''  issue!)\ the coefficient of this term in
$T(t)$ actually turns out to be zero.  However, recent
work in physics indicates that the $T$ expansion does not
give a trustworthy model of the energy in a realistic
system, and in a better model a nonzero $O(t^{-1})$ 
contribution reappears (rendered finite by regularization).
The paper works through these observations, roughly in order.

\section{Mathematical setting and notation}  \label{sec:notat}
Although the setting could  be greatly generalized,
here we assume that 
$\Omega$ is a compact region in $\mathbb{R}^{d}$ with smooth 
boundary, and
$H$ is an associated positive self-adjoint operator 
with pure point spectrum.
In the situation of most interest for Casimir physics,
$-H={\nabla}^{2}$ is the Laplacian on scalar functions 
in $\Omega$ with the Dirichlet boundary condition.
We use the two notations
\beq
{\lambda}_{n}={\omega}_{n}^{2}
\label{eigenvalues}\eeq
for the $n$th eigenvalue of $H$.

The heat kernel, $K(t,\mathbf{x},\mathbf{y})$, 
solves the initial-value problem for 
$-\frac{\partial u}{\partial t}=Hu$:
\beq
u(t,\mathbf{x})=e^{-tH}u(0,\mathbf{x})=
\int_\Omega K(t,\mathbf{x},\mathbf{y}) f(\mathbf{y})\,
  d\mathbf{y}.
\label{heatkernel}\eeq
We have the famous heat kernel expansion,
\beq
K(  t,\mathbf{x},\mathbf{y})  \sim( 4\pi t)^{-d/2}
e^{{|\mathbf{x}-\mathbf{y}| }^{2}/4t}
\sum_{s=0}^{\infty}a_{s}(\mathbf{x},\mathbf{y})t^{s/2},
\label{heatseries}\eeq
\beq
\Tr K=\int_{\Omega} K(  t,\mathbf{x},\mathbf{x})\, d\mathbf{x}
  \sim(  4\pi t)^{-d/2}\sum_{s=0}^{\infty}a_{s}[\Omega]t^{s/2}.
\label{heattrace}\eeq

Less well known, but more pertinent to vacuum energy, is the
cylinder (or Poisson) kernel, $T(t,\mathbf{x},\mathbf{y})$.
It solves the initial-value problem for 
\beq
-\frac
{\partial{^{2}}u}{\partial t^{2}}=Hu, 
\quad\lim_{t\rightarrow+\infty}u(t,\mathbf{x})=0:
\label{cylproblem}\eeq
\beq
u(t,\mathbf{x})=e^{-t\sqrt{H}}u(0,\mathbf{x})
=\int_\Omega T(t,\mathbf{x},\mathbf{y}) f(\mathbf{y})\,
  d\mathbf{y}.
\label{cylkernel}\eeq
$T$ is the $t$-derivative of another kernel, $\tbar$,
which solves the same problem as \eref{cylproblem}
except that $\frac {\partial u(0,\mathbf{x})}{\partial t}$
  is the initial data.
(Note that $t$ in \eref{heatkernel} and \eref{cylkernel}
can be thought of as related by a Wick rotation 
to the physical time in nonrelativistic (Schr\"odinger)
and relativistic (wave) equations, respectively.)
The cylinder kernel has a trace expansion \eref{cyltrace}
  similar to, and related to,  \eref{heattrace},
  to which we turn in \sref{sec:cyl}.

Let 
$N(\lambda)$ be the number of eigenvalues less than or equal to
$\lambda$.
Then the density of eigenvalues --- the derivative of $N$ ---
  is a distribution, having a Dirac delta function at each
   eigenvalue.
Also, it depends on the variable of integration:
$
\frac{dN}{d\omega} \ne
\frac{dN} {d\lambda}\,.
$
Instead,
\beq
\frac{d N}{d\omega} =2\omega\, \frac{dN}{d(\omega^{2})} =
  2\sqrt{\lambda} \,\frac{d N}{d \lambda}\,.
\label{eigdensities}\eeq

Famously, the counting function $N$ obeys \emph{Weyl's law}:
As $\omega\to\infty$ ,
\beq
N(\omega^{2}) \propto\omega^{d}.
\label{weyllaw}\eeq
(This holds  for a second-order operator in dimension
$d$ acting in a \emph{compact} region $\Omega$.)

\section{Properties and problems of the Weyl series}
\label{sec:weylproblem}

An obvious question is whether \eref{weyllaw} is the 
start of an asymptotic expansion;
that is, whether one can write something like
\[
N(\omega^2)\sim\sum_{s=0}^{N}g_{s}\omega^{d-s}\,.\]
Because $\Tr K$ is the Laplace transform of $\frac{dN}{d\lambda}$,
one can show,
by calculations like those in \cite{lgacee1},
  that the coefficients in this series,
if it existed, would be determined by the %heat kernel 
coefficients $a_s[\Omega]$
  in the rigorous asymptotic series  \eref{heattrace}.
(The Laplace transform of $\lambda^{p-1}$ is proportional
to $t^{-p}$, at least for $p>0$.)
However, it turns out that
only the leading term of the Weyl series is genuinely
asymptotic:
\beq
N(\omega^{2})=\sum_{s=0}^{M}g_{s}\omega^{d-s}+E_{M}(\omega)
\label{weylapprox}\eeq
where $E_{M}(\omega)$ is usually \emph{not} of order 
$O(  \omega^{d-M-1})  $.
  Instead, when $M>0$,
  in general $E_{M}$ is of the same order as the previous terms
  in the series (but is oscillatory).
This problem has been understood for ages; the
oscillations are related to periodic orbits 
of the classical system with Hamiltonian~$H$.

A related but less well known problem is that 
the proposed series and its formal derivatives become
  quite problematical
when the exponents of $\omega$ cease to be positive.
Consider, for example, the term $g_d \,\omega^0$.
Its derivative vanishes, so the eigenvalue density
$\frac{dN}{d\lambda}$ cannot contain a term proportional
to $\lambda^{-1}$.  In fact, if  a term with that 
asymptotic behavior did exist,
its Laplace transform would contain  $\log t$,
and the same is true of any negative integral power of $\lambda$;
but one knows that such terms do not appear in the heat 
kernels of second-order differential operators.
Where, then, did the heat-kernel coefficient $a_0[\Omega]$ 
come from? 
It, and the  $a_s[\Omega]$ with $d-s$ both negative and even,
come entirely
from the contributions of small values of $\lambda$
to the Laplace transform integral,
not from the asymptotic behavior of $N$ at large $\lambda$.

Two approaches have been followed to bring clarity and
precision into this seemingly rather muddled situation:
\begin{itemize}

\item \emph{Riesz--Ces\`aro means:}
The heat-kernel coefficients can be related by the Laplace 
transformation
to the coefficients of the asymptotic expansions of 
sufficiently high-order iterated antiderivatives
of the counting function \cite{FG,systemat}.

\item\emph{Generalized Weyl series and distributional moments:}
The ``missing" coefficients 
(which ``should'' accompany negative integral powers of $\lambda$)
can be identified as those multiplying
terms $\delta^{(j)}(\lambda)$ in the asymptotic analysis of
$N(\lambda)$
\cite{dowNaples,KZLS}.
These are known as 
``moment terms" in the  theory of distributions
\cite{estCes,EGbV,EF,EKbook}.

\end{itemize}
One of our principal goals is to elaborate on the second of these.

\section{Distributions and moment expansions}
\label{sec:dist-mom}

\subsection{Review of general theory} \label{subsec:gendist}

Recall that distributions are defined as linear functionals
on spaces of very well-behaved functions called 
\emph{test functions}.
A convenient test-function space is
\beq
\mathcal{D}(  \mathbb{R}^{n})  =
\left\{  \phi\in C^{\infty}(
\mathbb{R}^{n}) 
\colon \phi\mbox{ compactly supported.}
\right\}
\label{testspace}\eeq
The corresponding space of distributions is the \emph{dual space}
$\mathcal{D}'(\mathbb{R}^{n})$, conprising
  the linear functionals on
$\mathcal{D}(  \mathbb{R}^{n})$ 
(that are continuous in the weak topology).
The action of $f\in \mathcal{D}'$ on $\phi\in\mathcal{D}$
is written $\langle f(\mathbf{x}),\phi(\mathbf{x})\rangle$
and generalizes the inner-product integral
\beq
\int_{\mathbb{R}^n} f(\mathbf{x})\phi(\mathbf{x}) \, d\mathbf{x},
\label{distintegral}\eeq
to which it reduces when $f$ is a nonsingular ordinary function.
These concepts extend to, for example, functions and distributions 
defined on $\Omega$, with technicalities we shall not discuss here.
Also, we sometimes need to refer to distributions taking
values in some other vector space 
(rather than $\mathbb R$ or $\mathbb C$),
as in the case of $D^{\prime}(  \mathbb{R},L
(  \mathcal{X},\mathcal{H}))  $
in Definition~\ref{def:specden}.

\emph{Moments} are defined as the results of applying distributions 
to power functions.
Because the powers do not have compact support, it is necessary
to work with a larger test-function space and hence a more
restricted distribution space.
Loosely speaking, we need the integral \eref{distintegral}
to converge when $\phi(x)= x^q$.
Let
\beq
\mathcal{K}(  \mathbb{R}^{n})  =
\bigl\{  \phi\in C^{\infty}(
\mathbb{R}^{n})  \colon \exists q\in\mathbb{N} \colon
D^{\mathbf{k}}\phi(  \mathbf{x})  =O
\bigl(  \left\vert \mathbf{x}\right\vert
^{q-\left\vert \mathbf{k}\right\vert }\bigr) 
\quad\mbox{as }\left\vert
\mathbf{x}\right\vert \rightarrow\infty  \bigr\}.  
\label{Kspace}\eeq
(More technically, 
one defines the space $\mathcal{K}^{[  q]  }$
of functions satisfying the condition with a fixed $q$,
which is a Frech\'et space under certain seminorms,
and then defines $\mathcal{K}$ as the union of those spaces
with the inductive limit topology.)

The dual space, $\mathcal {K}'(  \mathbb{R}^{n}) $, 
is the distribution 
space where moment asymptotic expansions work.
Crudely speaking, 
$f\in \mathcal {K}'(  \mathbb{R}^{1}) $ says that
$f$ falls off sufficiently fast at infinity that all the 
moments $\langle f(x), x^q \rangle$ exist \cite{estCes}.

\begin{theorem}[moment asymptotic expansion theorem]
{\rm \cite[Sec.\ 3.3]{EKbook}}
Function $f$ is in $\mathcal{K}^{\prime}(  \mathbb{R}^{n})  $
if and only if
  $f(\lambda\mathbf{x})  $
  admits the \emph{moment asymptotic expansion}:%
\begin{equation}
f(  \lambda\mathbf{x})  \sim
\sum^\infty_{| \mathbf{k}| =0}
\frac{\left(
-1\right)  ^{\left\vert \mathbf{k}\right\vert } 
\mu_{\mathbf{k}}D^{\mathbf{k}%
}\delta(  \mathbf{x})  }
{\lambda^{\left\vert \mathbf{k}\right\vert
+n}\mathbf{k}!}\quad \mbox{as }\lambda\rightarrow\infty.
\label{usualexpansion}%
\end{equation}
Here 
\beq
\mu_{\mathbf{k}}=\left\langle f(  \mathbf{x})
,\mathbf{x}^{\mathbf{k}}\right\rangle =
\left\langle f(  \mathbf{x})
  ,x_{1}^{k_{1}}...x_{n}^{k_{n}}\right\rangle,
\qquad \mathbf{k}\in \mathbb{N}^n .
\label{momentdef}\eeq
\end{theorem}

Formula \eref{usualexpansion} means that if $\phi\in\mathcal{K}
(  \mathbb{R}^{n})  $, then for each~$M$,
\[
\left\langle f(  \lambda\mathbf{x})  ,\phi(\mathbf{x})\right\rangle =
\sum^M_{| \mathbf{k}|=0}
\left\langle \frac{\left(  -1\right)  ^{\left\vert \mathbf{k}
\right\vert }\mu_{\mathbf{k}}D^{\mathbf{k}}\delta(  \mathbf{x})
}{\lambda^{\left\vert \mathbf{k}\right\vert +n}\mathbf{k}!},
\phi(\mathbf{x})  \right\rangle 
+o\left(  \frac{1}{\lambda^{M+n}}\right) 
\]
as $\lambda\rightarrow\infty$.
Notice that it is an expansion in the parameter $\lambda$, 
not in the variable $x$ (which will be $\omega$ or $\omega^2$ 
in our spectral applications).
It is this change in  point of view 
that makes it logically possible  to incorporate part of the
content of $f$ at small $x$ into the series formally describing
the behavior of $f$ at large $x$ \cite{dowNaples,KZLS}.
In one dimension, the theorem says that
$f\in\mathcal{K}^{\prime}(  \mathbb{R})  $ 
is equivalent to the applicability of the moment 
expansion
\begin{equation}
f(  \lambda\mathbf{x})  \sim
\sum^\infty_{k=0}\frac{(  -1)  ^{k}\mu_{k}\delta^{(k)}(  x)}
{\lambda^{k+1}k!}
\quad\mbox{as }\lambda
\rightarrow\infty,
\label{1dexpansion}\end{equation}
where $\mu_{k}=\left\langle f(  x)  ,x^{m}\right\rangle $.

\subsection{The spectral density}
\label{subsec:specdist}

Let $\mathcal{H}$ be a Hilbert space and let $H$ be a 
self-adjoint operator
  with domain $\mathcal{X}\subset \mathcal{H}$.
Then $H$ admits a spectral decomposition 
$\left\{  E_{\lambda}\right\}
_{\lambda=-\infty}^{\infty}\,$, such that (in a weak sense)
\[
H=\int_{-\infty}^{\infty}\lambda\, dE_{\lambda}\,.
\]
In most cases of interest in this paper, this equation is a
discrete sum 
\[
H = \sum_{n=1}^\infty \lambda_n P_n\,, 
\]
where $P_n = E_{\lambda_n} - E_{\lambda_{n-1}}$
is the orthogonal projection onto the eigenvectors with eigenvalue
$\lambda_n\,$.

\begin{definition} \label{def:specden}
The \emph{spectral density} $e_{\lambda}=dE_{\lambda}/d\lambda$ 
is the distribution in
$D^{\prime}(  \mathbb{R},L
(  \mathcal{X},\mathcal{H}))  $ such that
\[
\left\langle e_{\lambda},\phi(  \lambda)
  \right\rangle _{\lambda
}=\int_{-\infty}^{\infty}\phi(  \lambda)\,  dE_{\lambda}\,.
\]
\end{definition}
One sometimes denotes $e_{\lambda}$ by 
$\delta(  \lambda-H)  .$

For example, the identity operator is
$
I=\left\langle e_{\lambda},1\right\rangle
  =\int_{-\infty}^{\infty} dE_{\lambda}\,$,
and $H$ itself is
$
H=\left\langle e_{\lambda},\lambda\right\rangle 
=\int_{-\infty}^{\infty
}\lambda\, dE_{\lambda}\,,
$
meaning that 
\[
\left(  Hx|y\right)  =\left(  \left\langle e_{\lambda},
\lambda
\right\rangle _{\lambda}x|y\right)  =
\int_{-\infty}^{\infty}\lambda\, d\left(
E_{\lambda}x|y\right) \mbox{ for all $x\in\mathcal{X}$ and
$y\in\mathcal{H}$}.
\]
(Here we write the inner product as $\left( \cdot|\cdot\right)$
to avoid confusion with the evaluation of a distribution,
$\langle \cdot , \cdot\rangle$.)

Let $\mathcal{X}_{n}$ denote the domain of $H^{n}$ 
and let $\mathcal{X}_{\infty}=
\cap_{n=1}^\infty
\mathcal{X}_{n}$ .
Then
$
\left\langle e_{\lambda},\lambda^{n}\right\rangle
  =H^{n}\in L\left(
\mathcal{X}_{\infty},\mathcal{H}\right)  \mbox{ exists}.
$
Hence
$
e_{\lambda}\in\mathcal{K}^{\prime}(  \mathbb{R},
L(  \mathcal{X}_\infty,\mathcal{H})  )  .
$

\begin{theorem} {\rm \cite[(6.328)]{EKbook}}
The moment asymptotic expansion of $e_{\lambda\sigma}$ is
\beq
e_{\lambda\sigma}\sim
\sum^\infty_{k=0}\frac{\left(
-1\right)  ^{k}H^{k}\delta^{\left(  k\right)  }
(  \lambda)
}{\sigma^{k+1}k!}\quad \mbox{as }\sigma\rightarrow\infty.
\label{emom}\eeq
\end{theorem}

We are now ready to address the Weyl expansion specifically.
The trace of $P_n$ is the number of eigenvectors with 
eigenvalue~$\lambda_n\,$,
so the trace of $E_{\lambda}$ is $N(\lambda)$;
\beq
N(  x)     =
\sum_{\lambda_{n}\leq x}1
  =\sum_{n}\theta(  x-\lambda_{n}).
\eeq
Observe that
\beq
N^{\prime}(  x)  =
\sum_{\lambda_{n}\leq x}\delta(x-\lambda_{n})  ,
\eeq
the trace of $e_\lambda\,$.
However, if we integrate \eref{emom} term by term, we get a wrong 
answer, inconsistent with \eref{weyllaw}.
The problem is that 
when we take the trace of $E_{\lambda}$ to get $N(\lambda)$, 
the result is no
longer in $\mathcal{K}^{\prime}$,
so the moment expansion theorem does not apply without 
modification.
The (Ces\`aro-averaged)
asymptotic expansion 
(see \cite{estCes} and \cite[Chap.\ 6]{EKbook})
of $N(\lambda)$ contains not only the
  moment terms but
also powers of $\lambda$. 
The exponents may be positive or negative, but
negative integers do not occur (being replaced by the moments). 
Terms of
negative half-integer order become literally asymptotic only 
after repeated
indefinite integration to form the Riesz means (see
\sref{sec:riesz} and \cite{FG,systemat,funorman}).

In the Riesz means of sufficiently high order, 
the difference between moments
and power terms is washed out.
  Correspondingly, in $\Tr K$ there is no deep
distinction between integral and half-integral powers of $t$. 
Taking the
direct product with a one-dimensional system 
(the interval or the one-torus,
with $K\propto t^{-\frac12}$) interchanges odd and even powers 
of $\sqrt{t}$.

\subsection{Green kernels associated with spectral decompositions}
\label{subsec:greendist}

The same ideas apply to the Green functions, or integral kernels,
associated with operators.
An operator $H$ on $\mathcal{D}(\Omega )  $ 
can be realized by a distributional kernel
  $h\in
\mathcal{D}^{\prime}(  \Omega\times \Omega)  :$
\[
Hf\left(  x\right)  =\left\langle h(  x,y)  ,\,
f(  y)\right\rangle _{y}\,.
\]
For example,
\[
\left\langle \delta(  x-y)  ,\, f(  y)
\right\rangle =If\left(  x\right) ,
\]
\[
\left\langle H\delta\left(  x-y\right)  ,\, f
\left(  y\right)
\right\rangle =\left\langle \delta(  x-y)  ,
\, Hf\left(y\right)  \right\rangle =Hf\left(  x\right) .
\]
% end of example

Similarly, $e_{\lambda}$ has an associated kernel 
$e(  x,y;\lambda)  \in\mathcal{K}^{\prime}(  \mathbb{R},
\mathcal{D}^{\prime}(
\Omega\times \Omega) )  ,$ such that
\beq
\left\langle \left\langle e(  x,y;\lambda)  ,\,
f(  y)
\right\rangle _{y},\,\phi(  \lambda) 
\right\rangle _{\lambda} 
=\left\langle e_{\lambda}\,,\,\phi(  \lambda)
  \right\rangle _{\lambda}f\left(  x\right)
   =\phi(  H)  f(  x)  .
\label{spectralkernel}\eeq
Observe that
$
\left\langle e(  x,y;\lambda)  ,
\lambda^{n}\right\rangle _{\lambda
}=H^{n}\delta\left(  x-y\right).
$
So we have

\begin{corollary} {\rm \cite[(61)]{EF}}
\beq
e(  x,y;\sigma\lambda)  \sim 
\sum^\infty_{k=0}\frac{\left(  -1\right)  ^{k}H^{k}\delta\left(
x-y\right)  \delta^{\left(  k\right)  }(  \lambda)  }
{\sigma
^{k+1}k!}\quad \mbox{as }\sigma\rightarrow\infty.
\label{speckermoment}\eeq
\end{corollary}

Let $K (  t,x,y)  =\left\langle e(  x,y;\lambda),
\,e^{-\lambda t}\right\rangle $
  be the heat kernel, as introduced in \eref{heatkernel}.
Then
\begin{eqnarray}
\left\langle e(  x,y;\lambda)  ,\,
e^{-\lambda t}%
\right\rangle  &  =\frac{1}{t}\left\langle 
e(  x,y;\lambda/t)
,\,e^{-\lambda}\right\rangle \nonumber\\
&  \sim
\sum^\infty_{k=0}\left\langle \frac{\left(
-1\right)  ^{k}H^{k}\delta\left(  x-y\right)
  \delta^{\left(  k\right)
}(  \lambda)  t^{k}}{k!},\,e^{-\lambda}\right\rangle
\nonumber \\
&  =\sum^\infty_{k=0}\frac{\left(  -1\right)  ^{k}%
H^{k}\delta\left(  x-y\right)  t^{k}}{k!}\quad
\mbox{as } t\rightarrow0.
\end{eqnarray}
This is a representation of $K$ as a distribution on 
$\Omega \times \Omega$, so it does not directly give the
asymptotic expansion ``on diagonal''\negthinspace, 
\eref{heattrace}.

\section{Example:  The Dirac comb} \label{sec:diracexamp}

Kolomeisky et al.\ \cite{KZLS}  work out some examples
  where the moment
terms can be calculated by the Euler--Maclaurin formula,
because the eigenvalues are equally spaced (possibly after
a change of variable).
Here we study the simplest such case and reproduce the
conclusions of \cite{KZLS} by another route.

\begin{theorem}\label{thm:combmoments}
{\rm \cite[Lemma 2.11]{EGbV}}
If $g\in\mathcal{K}$ 
and if $\int_{0}^{\infty}g(  x)  dx$ 
is defined, then
\begin{eqnarray}
\sum_{n=1}^\infty g(  n\varepsilon)  &
=\left\langle\sum^\infty_{n=1}\delta(  x-n)
,\,g(  \varepsilon x)  \right\rangle \nonumber\\
&  =\frac{1}{\varepsilon}\int_{0}^{\infty}g(  x)\,
dx+\sum^\infty_{n=0}\frac{\zeta\left(  -n\right)
g^{\left(  n\right)  }(  0)  }{n!}\varepsilon^{n}
+o(
\varepsilon^{\infty})  ,
\label{comb1}\end{eqnarray}
where $\zeta(  x)  $ is the zeta function.
Thus $\zeta(  -n)$ are the moments in this case.
That is,
\beq
\sum_{n=1}^\infty\delta\left( \frac{x}{\varepsilon}-n\right)
\sim \theta(x)  +
\sum^\infty_{n=0}\frac{\left(
-1\right)  ^{n}\zeta(  -n)  \delta^{\left(  n\right)  }
(x)  }{n!}\varepsilon^{n+1}\quad\mbox{as }
\varepsilon\downarrow0.
\label{comb2}\eeq
\end{theorem}

The \emph{generalized Weyl expansion} found in \cite{KZLS} is
\beq
\sum_{n=1}^\infty g\left(  \frac{\pi n}{a}\right)
=\frac{\pi}{a}\int_{0}^{\infty}g(  q)\,  dq
-\frac{g(  0)}{2}-\frac{\pi g^{\prime}(  0)  }{12a}
+\frac{\pi^{3}g^{(3)  }(  0)  }{720a^{3}}-\cdots.
\label{genweyl1}\eeq
That is, if one defines
$G(  q)  =\sum\delta(  q-n)  ,$ 
then
\beq
G\left(  \frac{aq}{\pi}\right)  \sim\frac{a}{\pi}-
\frac{\delta(q)  }{2}+\frac{\pi\delta^{\prime}(  q)  }{12a}
-\frac{\pi
^{3}\delta^{\left(  3\right)  }(  q)  }{720a^{3}}
+\cdots\quad\mbox{as }a\rightarrow\infty.
\label{genweyl2}\eeq
  Here $a$
denotes the radius of the one-dimensional interval, 
half the distance between
the (Dirichlet) boundaries.

By  Theorem \ref{thm:combmoments}, 
if $g\in\mathcal{K}$ and if 
$\int_{0}^{\infty}g(  x)\,  dx$ is
defined, then
\beq
\sum_{n=1}^\infty g(  n\varepsilon)  \sim
\frac{1}{\varepsilon}\int_{0}^{\infty}g(  x) \, dx+
\sum^\infty_{n=0}\frac{\zeta(  -n)  g^{\left(  n\right)
}(  0)  }{n!}\varepsilon^{n}\quad
\mbox{as }%
\varepsilon\downarrow0.
\label{comb3}\eeq
If we let $\varepsilon=\frac{\pi}{a}$ and compute the 
zeta-function terms, we
obtain the same equation as \eref{genweyl1}.
Note that the first term is a Weyl (high-frequency asymptotic)
term and the rest are moment terms, related to the low-frequency 
content of $N^{\prime}$.

The most elementary application of this expansion is
when the eigenvalues are proportional to $n^2$.
Then
\[
N(x)  =\sum_{n\geq0} \theta(x
-n^{2}),   \qquad N^{\prime}(x)  =
\sum_{n\geq0}\delta( x-n^{2}).
\]
If we set $g(  x^{2})  =f(  x)  $, 
we can then calculate
\begin{eqnarray}
\left\langle \sum_{n=1}^\infty \delta(
x-n^{2})  ,\, g(  \varepsilon x)
   \right\rangle
&=\sum_{n=1}^\infty f(  \varepsilon^{1/2}n)
\nonumber\\
&\sim\frac{1}{2\varepsilon^{1/2}}\int_{0}^{\infty}
\frac{g(  x)
}{\sqrt{x}}dx\quad\mbox{as }\varepsilon\downarrow0
\label{nsqseries}\end{eqnarray}
for the first term. 
But, actually, the first term is all:
  For the moment terms we can calculate the moments of 
$N^{\prime}$ to be
\[
\mu_{k}   =\left\langle \sum_{n\geq0}\delta(x
-n^{2})  ,x^{k}\right\rangle
  =\zeta(  -2k)  =0.
\]
That is,
\beq
\sum_{n=1}^\infty \delta\left(  \frac{x}{\varepsilon}
-n^{2}\right)  =\frac{\theta(  x)}{  
2\varepsilon^{1/2}\sqrt{x}}
+o(  \varepsilon^{\infty})  \quad
\mbox{as }\varepsilon\downarrow0.
\label{Nmoments}\eeq

As an example of the example,
let $Hy=y^{\prime\prime}$ be considered on the domain 
$\mathcal{X}=\left\{  y\in
C^{2}\left[  0,\pi\right]  \colon y\left(  0\right)  =
y\left(  \pi\right)
=0\right\}  $ inside $L^{2}\left[  0,\pi\right]$  .
The eigenvalues are
$\lambda_{n}=n^{2}$, $n=1,2,3,\ldots$, 
with normalized eigenfunctions 
$\phi_{n}(  x)  =\sqrt{2/\pi}\sin nx$.
Therefore,
\[
e(  x,y;\lambda)  =\frac{2}{\pi}
\sum_{n=1}^\infty
\sin nx\sin ny\,\delta(  \lambda-n^{2}),
\]
and
\begin{eqnarray*}
e(  x,x,\lambda)   &  =\frac{2}{\pi}
\sum^\infty_{n=1}\sin^{2}nx\,\delta(  \lambda-n^{2})
  \\
&  =\frac{1}{\pi}\sum_{n=1}^\infty\left(  1-\cos
2nx\right)  \delta(  \lambda-n^{2}) .
\end{eqnarray*}
Hence
\beq
e(  x,x,\lambda/t)
    \sim\frac{1}{\pi}%
\frac{\theta(  \lambda)  t^{1/2}}
{2\sqrt{\lambda}}\quad\mbox{as }
t\downarrow0
\label{combmoms}\eeq
for $0<x<\pi$.
  From this we recover the asymptotics of the heat kernel:
\begin{eqnarray}
\left\langle e(  x,x;\lambda)  ,
e^{-\lambda t}\right\rangle  &
\sim\frac{1}{2\pi t^{1/2}}\int_{0}^{\infty}
\frac{e^{-\lambda}}{\sqrt{\lambda}}\,d\lambda\nonumber\\
&  =\frac{1}{\left(  4\pi t\right)  ^{1/2}}
\quad\mbox{as }t\downarrow0.
\end{eqnarray}
This result is nonuniform in $x$, so it does not give
the correct trace expansion \eref{heattrace}, which contains
an additional term representing the effect of the Dirichlet
(or alternative) boundaries.
% end of example

Furthermore, we can study the relation to the expansion of the 
cylinder kernel.
We know from \eref{Nmoments} that
\[
\sum^\infty_{n=1}\delta\left(\frac{x}{\varepsilon}-n^{2}\right) 
 =\frac{\theta(  x)  }
{2\varepsilon^{1/2}\sqrt{x}%
}+o(  \varepsilon^{\infty})  \quad
\mbox{as }\varepsilon
\downarrow0.
\]
If we set $x=\omega^{2}$,   we might expect
\[
\sum_{n=1}^\infty 
\delta\left(\frac{\omega^2}{\varepsilon}-n^{2}\right) 
 =\frac{\theta(  \omega)  }
{2\varepsilon^{1/2}\omega
}+o(  \varepsilon^{\infty})  \quad
\mbox{as }\varepsilon
\downarrow0.
\]
But in fact, it's not so trivial:
\begin{equation}
\sum_{n=1}^\infty\delta(  \omega^{2}-n^{2})
=\sum_{n=1}^\infty\frac{1}{2n}\left[  \delta(
\omega -n)  +\delta(\omega +n)  \right]  ,
\label{deltasq}\end{equation}
so by \eref{comb2} we have
\begin{eqnarray}
\sum_{n=1}^\infty 
\delta\left(\frac{\omega^{2}}{\varepsilon}-n^{2}\right)  
\sim\frac{\theta(  \omega)  }
{2\varepsilon^{1/2}\omega}
+\sum_{n=0}^\infty &\frac{\left(  -1\right)  ^{n+1}%
\zeta(  -n)  \delta^{(  n+1)  }
(  \omega)
}{2\left(  n+1\right)  !}\varepsilon^{\frac{n+1}{2}}
\nonumber \\
&\mbox{\hskip 1in}+o(  \varepsilon
^{\infty})  \quad \mbox{as }\varepsilon\downarrow0.
\label{deltasqmom}\end{eqnarray}
The significance of the extra terms in \eref{deltasqmom}
will become clear in the next two sections.

\section{The cylinder kernel} \label{sec:cyl}

Because the material of this section and the next has been
extensively covered before \cite{FG,systemat,funorman},
we shall be relatively brief.
But to atone for a certain vagueness at certain points in earlier
sections, we give complete and precise formulas.
Please consult those papers for references to the fundamental
work of Hardy and H\"ormander on which the theorems are based.

The cylinder kernel of \eref{cylkernel} has the trace expansion
\beq
\sum_{n=1}^{\infty}e^{-t\omega_{n}} = \Tr T 
= \sum_{s=0}^{\infty}e_{s} t^{-d+s}
+ \sum_{
{\scriptstyle s=d+1\atop \scriptstyle s-d 
\mbox{ \scriptsize odd}}
}^{\infty}f_{s} t^{-d+s} \log t.
\label{cyltrace}\eeq
It is convenient to redefine the expansion coefficients in the 
heat trace   \eref{heattrace} by
\beq
\sum_{n-1}^{\infty}e^{-t\omega_{n}{}\!^{2}} = \Tr K 
= \sum_{s=0}^{\infty}b_{s}
t^{(-d+s)/2} 
\label{heattraceb}\eeq
when treating its relations with the cylinder trace and with 
Riesz means of $N$.

\begin{theorem}
The coefficients in the cylinder and heat expansions are 
related by
\[
e_{s} = \pi^{-1/2} 2^{d-s} \Gamma\left(  \frac{d-s+1}2\right)
  b_{s}
\quad\mbox{if $d-s$ is even or positive},
\]
whereas if $d-s$ is odd and negative,
\[
f_{s} = \frac{(-1)^{(s-d+1)/2} 2^{d-s+1}}{\sqrt{\pi} \Gamma
\bigl((s-d+1)/2\bigr)}\,b_{s}\,, \quad
\mbox{but $e_{s}$ is undetermined by the
$b_{r}\,$.}%
\]
\end{theorem}

The new coefficients $e_{s}$ (with $d-s$ odd and negative) are 
\emph{new} spectral invariants. 
They are \emph{nonlocal} in their dependence on the
geometry of $\Omega$.
The first one, $e_{d+1}\,$, has the interpretation of renormalized Casimir
vacuum energy in quantum field theory.

\section{Riesz means, old and new}\label{sec:riesz}

Riesz means are a generic tool of long standing,
but we consider only those of the counting function, $N$.
The ``old'' Riesz means (with respect to $\lambda$) are defined by the
$\alpha$-fold iterated (simplex) indefinite integration:
\[
R_{\lambda}^{\alpha}N\,(\lambda)=\frac{1}{\alpha!}\,
\lambda^{-\alpha}%
\int^{\lambda}
\buildrel \alpha \over \cdots
\int N(\tilde{\lambda})\,d\tilde
{\lambda}.
\]
But we may also have Riesz means with respect to
$\omega=\sqrt{\lambda}$:
\[
R_{\omega}^{\alpha}N\,(\omega)=\frac{1}{\alpha!}\,
\omega^{-\alpha}\int^{\omega}
\buildrel\alpha\over\cdots
\int N(\tilde{\omega}^{2})\,d\tilde{\omega}.
\]

\begin{theorem}
There exist asymptotic formulas of the forms
\[
R_{\lambda}^{\alpha}N=\int_{s=0}^{\alpha}a_{\alpha s}
\lambda^{(d-s)/2}%
+O\left(  \lambda^{(d-\alpha-1)/2}\right)  ,
\]%
\[
R_{\omega}^{\alpha}N=\int_{s=0}^{\alpha}c_{\alpha s}\omega^{d-s}
+\sum_{
{\scriptstyle s=d+1\atop\scriptstyle s-d
\mbox{ \scriptsize odd}}
}^{\alpha}d_{\alpha s}\omega^{d-s}\log\omega
+O\left(  \omega^{d-\alpha-1}\log\omega\right)  .
\]
\end{theorem}

Contrast \eref{weylapprox}.  In words, the oscillations 
that prevent the Weyl series from being asymptotic beyond
the first term
are averaged out by the integrations, so that the corresponding
series for a Riesz mean is a valid asymptotic approximation
to a certain higher order.

Only the coefficients $a_{ss}$ are truly important; the
$a_{\alpha s}$ with $s<\alpha$ contain redundant information.

\begin{theorem} 
The heat-kernel coefficients are proportional to the old 
Riesz means:
\[
b_{s}=\frac{\Gamma\left(  (d+s)/2+1\right)  }
{\Gamma(s+1)}\,a_{ss}\,.
\]
The cylinder-kernel coefficients are related to the new Riesz 
means by
\[
e_{s}=\frac{\Gamma(d+1)}{\Gamma(s+1)}\, c_{ss} \quad
\mbox{if $d-s$ is even or
positive,}
\]
\[
f_{s}=-\,\frac{\Gamma(d+1)}{\Gamma(s+1)}\, d_{ss}\,, \quad e_{s}
= \frac
{\Gamma(d+1)}{\Gamma(s+1)}\, [e_{ss}+ \psi(d+1)d_{ss}]
\]
if $d-s$ is odd and negative.
\end{theorem}

It is therefore no surprise that the asymptotic coefficients of 
the old and
new Riesz means are related by formulas \cite{FG} very similar to 
those relating the
heat and cylinder coefficients. 
In particular, when $d-s$ is odd and negative,
$c_{ss}$ is \emph{undetermined} by the $a_{rr}\,$.
There are integral operations leading from old Riesz means to 
new Riesz means and vice versa:
\begin{itemize}
\item When going old $\to$ new, the new $c_{ss}$ arise 
from the lower limit of
integration, bringing in new information about 
$N(\omega^{2})$ at low frequencies.

\item When going new $\to$ old, the $c_{ss}$ are multiplied by 
numerical
coefficients that turn out to equal $0$ when 
$d-s$ is odd and negative, so
their information is lost in those cases.
\end{itemize}

\section{One very special term} \label{sec:kacterm}

The term of order $t^{0}$ in a heat kernel is geometrically 
dimensionless and
has topological significance. 
It counts eigenfunctions with eigenvalue zero.

  More precisely, when $H_{1} = A^{*}A$ and $H_{2} = AA^{*}$,
\[
\Tr K[H_{1}] - \Tr K[H_{2}]
\]
is independent of $t$ (only the $O(t^{0})$ terms fail to
cancel) and equals the \emph{index} of the operator $A$. 
Kolomeisky et al.\ \cite{KZLS} call it the \emph{Kac term}.
We propose that \emph{index term} is more descriptive.

The index term in $K$ corresponds to a constant term in 
(the averaged) $N$. More
precisely, since $N(\lambda)=0$ for $\lambda<0$, 
the term is a multiple of the
Heaviside function, $\theta(\lambda)$. 
Therefore, it gives rise to a multiple of the Dirac delta
distribution in the eigenvalue density $dN/d\lambda$ or 
$dN/d\omega$\,. Thus,
whether this term is a moment or a high-frequency asymptotic 
term depends on
which function one is looking at.

Because there is no $\,\log\omega$ term in $N$, 
there is no $O(\omega^{-1})$
term in $dN/d\omega$. Therefore, it is correctly said that 
the index term comes
from the low-frequency, not high-frequency, 
behavior of the spectral density.
Note, however, that it is a true heat-kernel term, 
locally determined by the
geometry, not one of the new cylinder-kernel terms.

The index term corresponds to a term $O(t^{0})$ in the cylinder 
trace, $T$.
There is no $O(\log t)$ term. Hence there is no 
$O(t^{-1})$ term in $dT/dt$.
$dT/dt$ has the physical interpretation of ($-2$ times) the
  vacuum energy of a
scalar field, subjected to 
\emph{timelike point-splitting regularization}
(see references in \cite{EFM,sector1}).
Negative powers of $t$ here represent buildups of energy 
against the idealized
boundary, which must be absorbed into the description of 
the boundary itself
(\emph{renormalization}).

We see that no renormalization is needed in order $t^{-1}$; 
nevertheless,
calculations of local energy density show that large boundary 
energy is indeed there!
(Energy density proportional to $ z^{-2}$ 
($z=\mbox{distance to boundary}$) formally
implies energy proportional to $t^{-1}$.)

Candelas \cite{cand} correctly stresses the physical 
necessity of the $O(z^{-2})$ energy
density and argues (apparently incorrectly) that an 
$O(t^{-1})$ term in the
total energy arises from $dN/d\omega$.
  Other authors \cite{BGH,KZLS} correctly point out
that no such term exists.
Here we attempt to dispose of this controversy.

First, in timelike point-splitting regularization the large 
energy density
near the boundary is compensated by a larger opposite-sign 
density
concentrated even closer to the boundary. 
Therefore, there is no mathematical contradiction
between the local and global statements.

Second, there are now physical reasons to believe that for
  certain purposes,
spacelike point-splitting gives a more trustworthy model 
of the energy in a
realistic system \cite{EFM,sector1}. 
In that framework the contribution of the index term does not
vanish after all.

Third, quite apart from the technical criticisms of \cite{cand}
in \cite{KZLS}, a close examination of \cite{cand} shows
that its argument (as concerns the total energy, not the boundary
energy density) does not note the possibility that the
overall coefficient of the term in question might turn out to 
be zero.  In fact, as just remarked, it \emph{is} zero
in timelike but not in spacelike point-splitting.

Thus a 30-year-old controversy has been revived, resolved, 
and rendered
irrelevant in roughly the same year (2012--13).

In summary, the index term has a number of special
properties, which are not particularly closely related to 
each other:
\begin{itemize}
\item It does not contribute to the total energy in timelike
  regularization,
because its contribution to $\Tr T$ is killed by the
  differentiation.

\item Because it sits on the boundary between moments and
  high-frequency
asymptotics, there is no corresponding $O(\lambda^{-1})$ 
term in the
eigenvalue density.

\item Some moments (low-frequency contributions) in $\Tr T$ 
are new spectral
invariants that do not appear in $\Tr K$. 
\emph{However, this term is not one of them.} 
They start immediately afterwards.
\end{itemize}
There is a tendency for these three issues to become muddled 
together in our thinking.

\section{Summary of the subtleties} \label{sec:summary}

\begin{itemize}
\item Certain powers in the heat-kernel expansion match moment (delta
function) terms (not powers) in the ``generalized Weyl expansion''.
 A generalized Weyl expansion that is introduced in the paper 
\cite{KZLS} can
be realized as the moment asymptotic expansion of the Dirac comb 
$%
\sum_{n=1}^\infty\delta\left( x/\varepsilon -n\right)$ acting 
on a test function.

\item The missing powers match terms in the cylinder-kernel 
expansion. 
These new, nonlocal spectral invariants show up in the 
Riesz-Ces\`aro asymptotics of
$N$ with respect to $\omega$ (but not $\omega^{2}$).

\item Some confusion and controversy in the physics literature 
is related to this fact: 
The term in the $t$-derivative of the cylinder kernel trace
corresponding to the ``index'' term in the heat kernel trace 
vanishes because of an algebraic accident, 
but nevertheless quantum field theory predicts a
divergent boundary energy density proportional to that spectral
  invariant.
That this term is a moment, 
not a high-frequency part of the eigenvalue
density, is beside the point. 
Recent physics suggests that this method of
regularizing the energy is not reliable anyway.
\end{itemize}

\ack
This research was supported by National Science Foundation Grants
PHY-0968269 and PHY-0968448.
Y. Yang thanks the Texas A\&M Mathematics Department for
  hospitality while much of the work was done.
  Its dissemination was aided by invitations to participate
  in the Special Session on Topics in Spectral Geometry and 
Global Analysis, AMS Western Section Meeting, Albuquerque,
and in the Casimir Physics Session, \'Ecole de Physique, Les 
Houches.
   We both thank  our mentor in distribution theory,
Ricardo Estrada.

\end{document}